\documentclass[letterpaper]{article} 
\usepackage[]{aaai23}  
\usepackage{times}  
\usepackage{helvet}  
\usepackage{courier}  
\usepackage[hyphens]{url}  
\usepackage{graphicx} 
\usepackage{natbib}  
\usepackage{caption} 
\usepackage{algorithm}
\usepackage{algorithmic}

\usepackage{newfloat}
\usepackage{listings}
\DeclareCaptionStyle{ruled}{labelfont=normalfont,labelsep=colon,strut=off} 
\floatstyle{ruled}
\newfloat{listing}{tb}{lst}{}
\floatname{listing}{Listing}
\usepackage{enumitem}  

\nocopyright

\title{The AI Triplet: Computational, Conceptual, and Mathematical\\ Knowledge in AI Education}
\author{
    Maithilee Kunda
}
\affiliations{
    Computer Science, Vanderbilt University, Nashville, TN USA \{mkunda@vanderbilt.edu\}\\

}

\usepackage{bibentry}

\begin{document}

\maketitle

\begin{abstract}
Efforts to enhance education and broaden participation in AI will benefit from a systematic understanding of the competencies underlying AI expertise.  In this paper, we observe that AI expertise requires integrating computational, conceptual, and mathematical knowledge and representations.  We call this the ``AI triplet,'' similar in spirit to the ``chemistry triplet'' that has heavily influenced the past four decades of chemistry education research.  We describe a theoretical foundation for this triplet and show how it maps onto two sample AI topics: tree search and gradient descent.  Finally, just as the chemistry triplet has impacted chemistry education in concrete ways, we suggest two initial hypotheses for how the AI triplet might impact AI education: 1) how we can help AI students gain proficiency in moving between the corners of the triplet; and 2) how all corners of the AI triplet highlight the need for supporting students' spatial cognitive skills.
\end{abstract}

\section{Introduction}

\begin{figure*}[t]
    \centering
    \includegraphics[width=\linewidth]{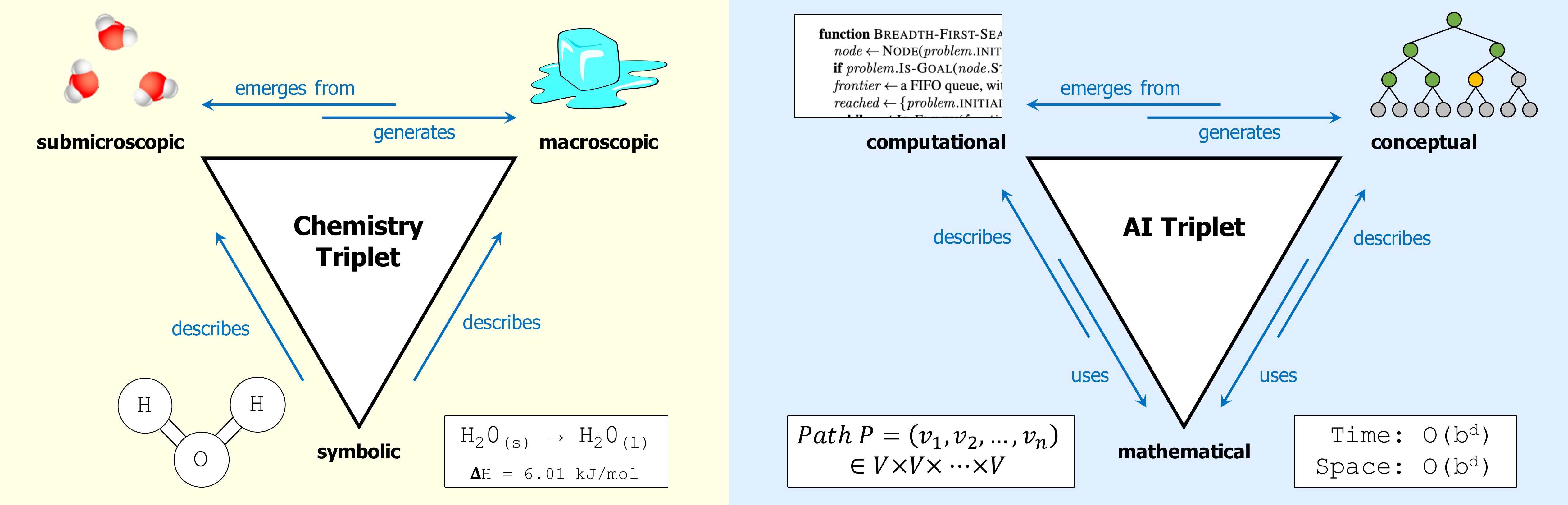}
    \caption{Illustrations of the chemistry triplet (left), and the proposed AI triplet (right).}
    \label{fig:bigfigure}
\end{figure*}

In 1982, a Scottish professor of chemistry and science education named Alex H. Johnstone published a tiny, 2.5-page paper in the \textit{School Science Review} that proposed the ``chemistry triplet'' as a new way of understanding the three types of knowledge required for chemistry expertise \cite{johnstone1982macro}.  Within the next 30 years, this little idea had grown to become ``one of the most powerful and productive ideas in chemical education'' \cite[p. 179]{talanquer2011macro} and ``a ‘taken-for-granted’ commitment (or assumption) for those working in the field'' \cite[p. 156]{taber2013revisiting}.

The chemistry triplet rests on a firm theoretical foundation of what chemistry is about, including its basic conceptual units as well as the role of human-created representations in how we formulate and communicate chemistry ideas.
Here, we propose an analogous AI triplet that we believe could be equally influential in the sphere of AI education.  Like Johnstone's chemistry triplet, the AI triplet rests on a firm theoretical foundation of what AI is about, including its basic conceptual units as well as the role of human-created representations in how we formulate and communicate AI ideas.

The time is exactly right for this kind of systematic examination of knowledge in AI.  As the use of AI is rapidly expanding into virtually all corners of the world, so too is AI education rapidly ballooning into more and more courses, dedicated degree programs, high school summer camps, and more, thus making research on AI education increasingly important.  Furthermore, AI is facing a crisis in its lack of diversity and inclusion, and the field desperately needs to improve educational practices as one part of addressing this problem.  We are optimistic that frameworks like the AI triplet can contribute to broadening participation in AI by providing new perspectives on pedagogical challenges and potential solutions for students who may not be well served by current approaches.
In this paper:
\begin{enumerate}[nolistsep,noitemsep]
    \item We make the case for our proposed AI triplet of computational, conceptual, and mathematical knowledge and representations.  We describe a theoretical foundation for the AI triplet as well as how it maps onto two sample AI topics: tree search and gradient descent.
    \item We describe how the AI triplet is compatible with but different in scope from Marr's three levels of analysis.
    \item We suggest two initial hypotheses for how, like the chemistry triplet, the AI triplet can inform insights and practices in AI education: 1) how we can help AI students gain proficiency in moving between the corners of the triplet; and 2) how all corners of the AI triplet highlight the need for supporting students' spatial cognitive skills.
\end{enumerate}




\section{A Brief Primer on the Chemistry Triplet}

Experts in chemistry integrate knowledge about chemical phenomena across three levels or types of knowledge: 
\cite{johnstone1982macro, gilbert2009introduction}:
\begin{itemize}[nolistsep,noitemsep]
    \item \textit{Macroscopic} refers to human-observable properties of substances, e.g., water turning from liquid to solid at a particular temperature.
    \item \textit{Submicroscopic} refers to properties at molecular and sub-molecular scales, e.g., the lattice shape formed by water molecules as they freeze, and the forces and geometry that cause this particular type of crystallization.
    \item \textit{Symbolic} refers to the mathematical, diagrammatic, and other notational formalisms of chemistry, e.g., the symbol $H_{2}O$ to indicate a water molecule, or the equation describing the change in energy when ice melts: $$H_{2}O (s) \rightarrow H_{2}O (l) \hspace{0.5cm} \Delta H = 6.01 kJ/mol$$
\end{itemize}

\noindent \textbf{Chemistry as a complex system.}  It is interesting to note that the first two levels---macro and submicro---describe not just how chemists think about chemistry but also, at a deeper theoretical level, chemistry itself.  In particular, chemical phenomena are an example of \textbf{complex systems} in nature.  

At the macro level, properties like temperature and viscosity describe high-level behaviors of substances.  These macro-level properties \textit{emerge} from the complex interactions of molecules at the submicroscopic level, which in turn emerge from yet lower levels of atomic or subatomic interactions \cite{luisi2002emergence}.  Going the other way, submicroscopic interactions can be said to \textit{generate} macro-level properties.  This relationship is illustrated in Figure \ref{fig:bigfigure}.

\textbf{The symbolic level.}  The first two levels---macro and submicro---are the portions of the chemistry triplet that actually ``exist'' in nature.  The third level, symbolic, refers to the human-created trappings of scientific notation used to \textit{describe} chemical phenomena at the other two levels, including chemical and mathematical symbols, equations, molecular diagrams, drawings, graphs, and so on \cite{taber2013revisiting}.


Some researchers subdivide the symbolic level depending on whether notation is used to describe mathematical calculations versus non-mathematical concepts \cite{nakhleh1994influence}.  For example, acids and bases are often described using chemical formulae, which are not themselves directly related to calculations, e.g., $H_2SO_4$ and $NaOH$, while other notation is used to describe equations explicitly used for calculation, such as the equation relating pH to the concentration of $H^+$ ions: $-pH = log[H^+]$.  

Other researchers distinguish between depictive versus non-depictive notation \cite{hoffmann1991representation}.  Depictive symbols can serve as \textit{iconic} models of underlying concepts, i.e., as representations that share some structural correspondence with what they represent, and thus provide a reasoning agent with additional inferential affordances regarding those concepts \cite{nersessian2010creating}.  For example, the chemical formula $H_2O$ describes the atomic contents of a water molecule, but the simple diagram $H{-}O{-}H$ provides additional information about the structure of a water molecule, as reflected in the structure of the diagram itself.

Depictive representations can reside in different corners of the chemistry triplet depending on particular use cases \cite[p. 184-185]{talanquer2011macro}:

\begin{quote}
``The semi-symbolic, semi-iconic nature of many visual representations in chemistry gives them a hybrid status between signs and models.... If we think of them as mere signs, then we may be inclined to classify them as belonging to the [symbolic] level; if we think of them as models with descriptive, explanatory, and predictive power we may prefer to think of them as part of the [submicro] level.''
\end{quote}

Following this logic, for our AI triplet, instead of separating out all symbolic notation into its own category, we instead consider notation as an additional layer of representation that resides above all parts of the triplet.  And, we give mathematical knowledge and notation special status in one corner of the triplet, as described in the following sections.



\section{The Proposed AI Triplet}

We propose the following AI triplet of types of knowledge:
\begin{itemize}[nolistsep,noitemsep]
    \item \textit{Computational} refers to the formal computations that make up programs together with the physical systems that run them, e.g., a computer running a program for depth-first search.
    \item \textit{Conceptual} refers to the abstract constructs that are accessed and manipulated (either explicitly or implicitly) via computations, e.g., the ``search tree'' that is being traversed by a program running depth-first search.
    \item \textit{Mathematical} refers to the mathematical formalisms used to either define or describe programs, e.g., using big-oh notation to describe the worst-case run time of depth-first search given a certain type of search tree.
\end{itemize}

\noindent The following subsections present a detailed rationale for each corner of this AI triplet, with relationships illustrated in Figure \ref{fig:bigfigure}.  The main observation motivating this formulation of the AI triplet is that, as with chemistry, \textbf{computer programs are examples of complex systems,} and part of computer science involves the empirical study of these systems \cite{newell1975computer}.  As Simon later wrote:

\begin{quote}
``An artificial system, like a natural one, produces empirical phenomena that can be studied by the methods of observation and experiment common to all science.  It might be objected that a system designed deliberately to behave in a desired way can produce no surprise or new information. This objection shrugs off our enormous ignorance of natural law and of the effects produced by natural laws operating on complex systems. The world of artificial (and natural) objects is full of unanticipated consequences, because of the limits both of empirical knowledge and of computational power.
'' \cite[p. 99]{simon1995artificial}
\end{quote}



In a nutshell, computations in AI (e.g., lines of code) are the ``submicro'' elements that interact to generate ``macro'' level behaviors at the conceptual level that can be observed, studied, and used by people.  Mathematics is used in AI in two primary ways: it is used within AI systems to \textit{define} things that are computed, and it is also used by people to \textit{describe} the operations or performance of such systems.  
And, as with chemistry \cite[p. 377]{johnstone1982macro}, AI experts ``jump freely from level to level in a series of mental gymnastics.  It is eventually very hard to separate these levels.''  

\subsection{Submicroscopic $\iff$ Computational}

If we draw an analogy between the complex natural systems of chemistry and the complex artificial systems of AI, then the low-level, causal elements in AI---i.e., the elements that make up the ``submicroscopic'' level of AI---are essentially lines of code: the pieces of formal computation that make up an artificial system or program.  Just as submicroscopic interactions are what \textit{generate} higher-level phenomena in chemistry, computation is what \textit{generates} higher-level phenomena in AI.

\begin{quote}
We define the \textbf{computational level} of the AI triplet as having to do with the computation that takes place in an artificial system, e.g., as specified by its program.
\end{quote}

Unlike chemistry, in which we are still working to understand submicroscopic processes, in AI, we know the exact rules that govern computation, and computations are directly observable.  However, as noted in Simon's quote above, the results of running a program are not always evident just from inspection of the code.  Higher-level behaviors emerge from the interactions of lower-level elements, and as with all complex systems, the higher-level behaviors cannot often or easily be predicted, even with full knowledge of the starting conditions and the rules of the system.  

The halting problem is one obvious example of this property.  Another ready example can be found in deep learning, in which slight changes in the initial conditions, i.e., hyperparameter settings, can drastically change the behavior of the final network \cite{feurer2019hyperparameter}.

Moreover, just as the submicroscopic level in chemistry can be broken into molecular, then atomic, and then subatomic interactions, so too can the computational level in AI be broken into human-readable code, then assembly code, then machine code, and so on.  
And, just as the submicroscopic level in chemistry eventually devolves into quantum physics, so too does the computational level in AI eventually devolve into electrical engineering (and eventually also into quantum physics!), assuming computations are realized on a digital electronic computer.  

However, it is still useful to describe these phenomena at higher levels of abstraction, e.g., at the level of molecules, for chemistry, and human-readable computations, for AI.  


\subsection{Macroscopic $\iff$ Conceptual}

In chemistry, the higher-level phenomena that emerge from low-level submicroscopic processes are easy to discern: they are the everyday, human-scale manifestations of matter that can be described by casual observers in terms of sensory impressions or more formally by experts in terms of properties like temperature and viscosity.

However, the higher-level phenomena emerging from low-level computations in an AI system are not so easy to discern.  
In a sense, the ``human-scale'' manifestations of executing an AI program are artificial constructs that exist in some abstract, conceptual space.  These constructs are often only partially built or accessed by the artificial system, and are also often only partially or vaguely represented in the mind of the human observing the system.  However, the constructs themselves transcend both of these partial views.  

For example, consider a program that performs depth-first search over a tree.  In many cases, the tree does not exist as an explicit entity in the program; it might be encoded implicitly as a starting node and a successor function.  In addition, we (as human observers) do not have explicit access to the complete tree in our own mental representations.  

However, 
the complete tree can still be defined as an abstract but fully-specified conceptual construct.  The complete tree ``exists'' in the same space that abstract mathematical objects might be said to ``exist,'' like, for instance, the set of all integers.  And, just as we can concretely use a few integers to count up a batch of jelly beans, partial views of an abstract search tree can be instantiated as concrete objects and used for some purpose, via the execution of a program.


\begin{quote}
    We define the \textbf{conceptual level} of the AI triplet as as having to do with the abstract conceptual constructs that are built and/or accessed by an artificial system.
\end{quote}


In an interesting reversal between the chemistry triplet and the AI triplet, the levels associated with \textit{generative} phenomena versus \textit{emergent} phenomena are inverted in terms of which levels are partially versus more fully observable.  

In the complex systems of chemistry, the submicroscopic level (which generates higher-level phenomena) is not directly observable by people, 
and we can often access only partial or indirect information about this level through specialized equipment, partial observations, experiments, and our imagination (e.g., mental models and thought experiments).  The macroscopic level (which represents the emergent phenomena in the complex system) is directly (though still not fully) observable by people.

In contrast, in the complex systems of AI, the computational level (which generates higher-level phenomena) is directly observable by people.  However, the conceptual level (which encompasses the emergent phenomena in the complex system) is not directly observable by people.  And, as with the submicroscopic level in chemistry, 
we can often access only partial or indirect information about this level through our experiments, observations, and imagination.


\subsection{Symbolic $\iff$ Mathematical}

First of all, it is worth pointing out that the notion of ``symbols'' in chemistry is vastly different from what we mean by ``symbols'' in AI.  For the remainder of this paper, we use the term ``symbolic'' as it is used in the context of the chemistry triplet, i.e., the use of signs \textit{by humans} to convey ideas about a scientific topic \cite{talanquer2011macro}.
There are many types of symbolic notation used in AI, including:
\begin{enumerate}[nolistsep,noitemsep]
    \item Code (and pseudocode).
    \item Visual diagrams of conceptual constructs such as trees, hyperplanes, neural networks, etc.
    \item Mathematical notation such as equations, big O notation for algorithmic complexity, etc.
    \item Visual plots of quantitative relationships, such as data plots or Cartesian graphs.
\end{enumerate}

\noindent As described above regarding the chemistry triplet, while Johnstone's original triplet lumped together all symbolic notation into one corner, more recent efforts have found value in (a) linking the representation of specific types of knowledge to their respective corners of the triplet \cite[e.g.,][]{talanquer2011macro}, and also (b) separating out mathematical concepts and notation as a specialized form of knowledge \cite[e.g.,][]{nakhleh1994influence}.  
Following similar logic, we consider various kinds of notation in AI as being potentially associated with different corners of the triplet, depending on what type of knowledge a particular piece of notation describes, and we assign the third corner of the triplet to deal specially with mathematical concepts and notation.  

Next, we briefly discuss each type of AI notation listed above and where it would reside in our AI triplet.


\textbf{(1) Code and pseudocode} are of course an essential part of AI notation.  These not only \textit{describe} phenomena at the computational level of the AI triplet, but code in fact \textit{defines} computations.  
Because code is so intimately tied to the phenomena of computation, we suggest that this type of notation should reside at the computational level of the AI triplet.

\textbf{(2) Visual diagrams} are often used to describe phenomena at the conceptual level of the AI triplet.  For example, we often draw downward-fanning trees to illustrate principles of tree search.  Diagrams of neural networks show the arrangement and connections of nodes.  These kinds of diagrams can usefully be deemed to reside at the conceptual level of the AI triplet.  Of course, other types of visual diagrams might reside at other corners of the AI triplet, but we argue that these conceptual diagrams in AI have special status, as they are often a primary means by which human experts represent and communicate ideas about abstract AI concepts like search trees.


\textbf{(3) Mathematical concepts and notation} are used extensively in AI, and we argue that mathematics embodies a distinct, third \textit{type} of knowledge that is essential in AI expertise.  We see two primarily roles for mathematics in AI.

First, artificial systems might be defined to perform mathematical calculations during their operation, e.g., a program for doing gradient descent will contain some calculus-based method for computing the derivative of a function.  We label this as an artificial system \textit{using} mathematics.

Second, we (the human observers) frequently rely on mathematics for describing artificial systems.  For instance, we often define the inputs and outputs of a system in terms of variables and/or certain types of mathematical objects.  

We also often use mathematics to describe the behavioral or operational characteristics of artificial systems.  
For example, given a tree with branching factor $b$ and depth $d$, we can use some combinatorics and our knowledge of search algorithm behavior to mathematically infer that breadth first search will have a worst-case space complexity of $O(b^d)$.  

Regardless of whether mathematics are used within an artificial system itself or by human observers while describing such a system, these mathematical ideas and notation are distinct from both the computational level and the conceptual level of the AI triplet.  For example, in a program for training a neural network, the sigmoid function does not exist as a generalized equation anywhere in the actual computations that are carried out by an artificial system; only specific instances of it will ever be computed while the system is running.  However, the sigmoid function exists as a distinct mathematical entity whose properties we leverage in the design of a neural network.  Similarly, 
while at the conceptual level, we might visualize a tree with branching factor $b$ and depth $d$, and at the computational level, we might understand that breadth-first search in the worst case might have to search the entire tree, it takes a bit of additional mathematical maneuvering to arrive at the total number of nodes in the tree as $b^d$.
Thus, we propose elevating mathematical representations to form the third leg of our proposed AI triplet:


\begin{quote}
We define the \textbf{mathematical level} of the AI triplet as having to do with the mathematical notation and ideas used to describe various aspects of an artificial system, including its internal procedures as well as externally observed characteristics.
\end{quote}


\textbf{(4) Visual plots} are ubiquitous in AI, as in all empirical sciences. 
As with mathematical ideas, visual plots in AI are sometimes used to describe an internal component of an AI system, such as the graph of a loss function used for training a neural network.  Other visual plots might reflect external observations about the system, for instance a neural network training curve of error over time.  Interpreting such plots often requires both mathematical \cite{lee2019correlation} and visuospatial \cite{kozhevnikov2002revising} proficiency.  Thus, we suggest that visual plots span both the conceptual and mathematical levels of the AI triplet.

\subsection{Sample AI topics: BFS and gradient descent}

\begin{figure*}[t]
    \centering
    \includegraphics[width=\linewidth,trim={0 0.35cm 0 0},clip]{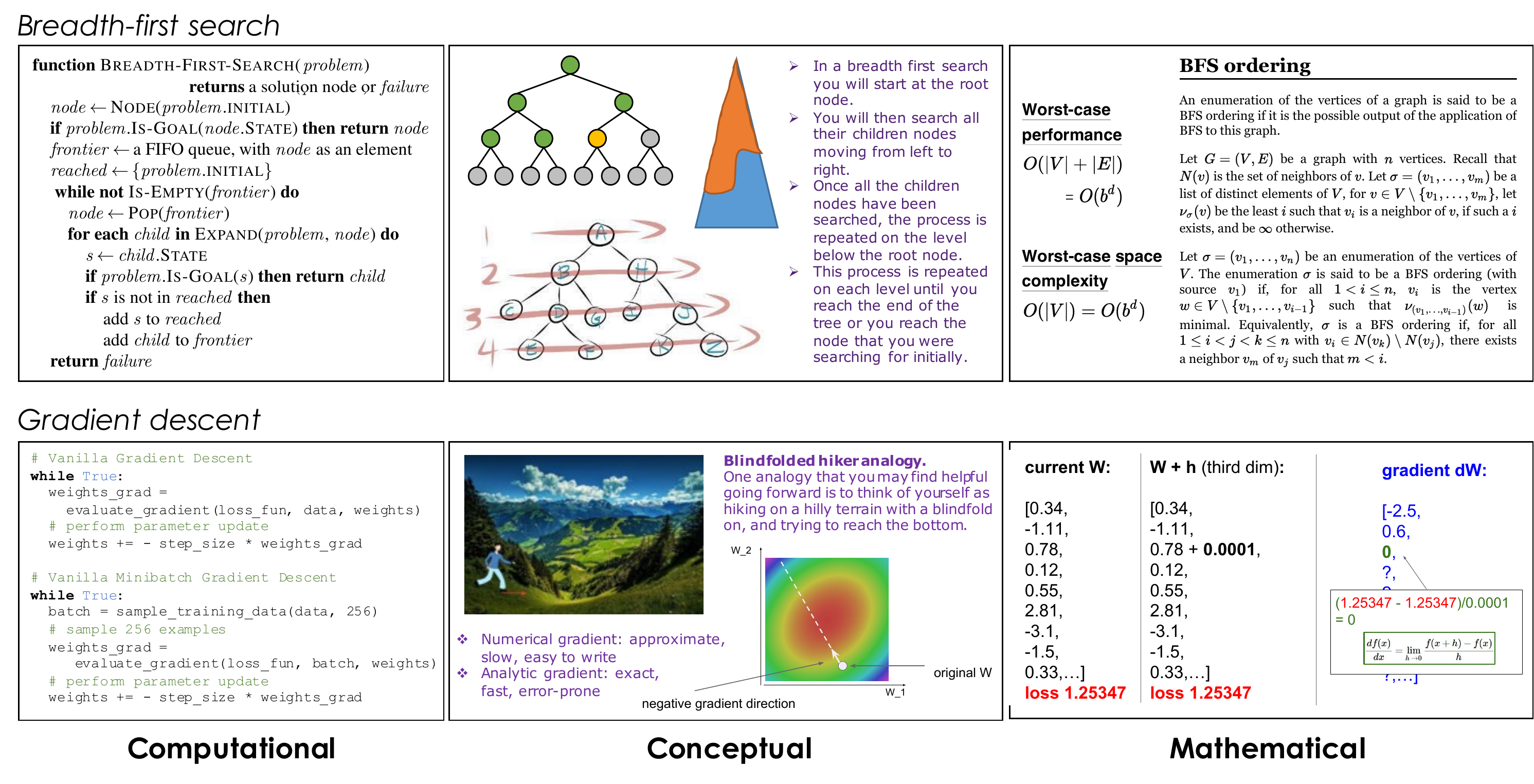}
    \caption{Sample AI topics with materials presented from all three corners of the AI triplet.  Top: Breadth-first search, with materials from \cite{russell2021artificial}, \cite{bland2017breadth}, and \cite{wiki:breadthfirstsearch}, respectively.  Bottom: Gradient descent, with materials from \cite{li2019optimization}.}
    \label{fig:bfs}
\end{figure*}

AI topics are commonly taught using material across all three corners of the AI triplet, though of course not always explicitly labeled as such.  Figure \ref{fig:bfs} shows publicly available teaching materials for two sample AI topics of breadth-first search (BFS) and gradient descent.

While it is not surprising that topics are taught with computational and mathematical content, the figure highlights how conceptual knowledge is often as prominent as the other two types of knowledge.  Conceptual knowledge often involves diagrams, analogies, and other types of informal illustrations, and often refers to abstract structures that underlie a particular AI technique, such as the notion of search trees or the surface of an optimization function.


\newpage
\section{Relationship to Marr's Levels of Analysis}

When considering this AI triplet, some readers may be thinking, ``We already have an AI triplet!  Isn't this just a rehash of Marr's three levels of analysis?''  While both frameworks do involve the number three, they actually represent different and orthogonal classifications of knowledge, i.e., they are different in scope, and not mutually exclusive.


Marr's three levels refer to different lenses or levels of abstraction through which to study an information processing system \cite{marr1976understanding}.  The levels often go by varying labels, but generally can be summarized as:
\begin{enumerate}[nolistsep,noitemsep]
    \item \textit{Computational, functional, or behavioral level:} The outwardly observable behavior that a system produces, for example, in terms of input-output mappings.
    \item \textit{Representational or algorithmic level:} The internal information-manipulation procedures that a system uses to produce its outward behaviors.
    \item \textit{Implementational or hardware level:} The physical substrate on which the information processing system is run.
\end{enumerate}

\noindent For example, let us use Marr's levels to analyze an AI system for search.  The computational level is analogous to the program's function header: the program might take as input a starting point, goal, and successor function, and return as output a path to the goal, if one exists.  
The representational level is analogous to the function body, including all of the intermediate data structures, operations, etc., needed to produce the input-output behavior, e.g., a priority queue to store nodes during search, the actual search procedures, etc.
Finally, the implementational level refers to the actual physical machine that our search program is running on.


Marr's levels essentially carve up information processing systems into horizontal slices.  At the top level, you have the input-output behavior.  This rests on the middle level of representations and algorithms.  And this middle level rests on the bottom layer of physical implementation.

Our AI triplet, in contrast, carves up information processing systems into vertical slices.  Any of Marr's 3 levels can be seen in any corner of our AI triplet, and vice versa.
For example, keeping strictly to Marr's level 1, the input-output behavior of a search program can be viewed in terms of:
\begin{enumerate}[nolistsep,noitemsep]
    \item \textit{(AI triplet level 1)} Its computational definition, e.g., input arguments and parameters and return values.
    \item \textit{(AI triplet level 2)} Its conceptual meaning and abstract structures, e.g., search trees, connected paths, etc.
    \item \textit{(AI triplet level 3)} Relevant mathematical formalisms, e.g., we can describe the output variable as a path $P$ across a set of graph vertices $V$, where $P$ is defined as: $P = (v_1, v_2, ..., v_n) \in V \times V \times ... \times V$.
\end{enumerate}


\noindent Similar examples can be constructed for Marr's level 2, e.g., we can describe specific search algorithms in terms of their computational, conceptual, and mathematical aspects.  

And likewise for Marr's level 3, in describing computational, conceptual, and mathematical aspects of a physically implemented computing system.

To summarize: \textbf{Marr's framework is about \textit{different levels of abstraction}} at which we can describe and analyze a given information processing system.  

\textbf{The AI triplet is about \textit{different kinds of knowledge and representations}} that we can bring to bear to describe one or more levels of abstraction in a given system.

\section{Discussion: The AI Triplet and AI Education}

If we consider the AI triplet as a way of organizing the types of knowledge needed for AI expertise, how can that help advance research and practices in AI education?  The chemistry triplet has catalyzed numerous concrete proposals for improving chemistry education, and we expect that the same could be true of AI education using the AI triplet.

Next, we discuss two initial examples of the kinds of hypotheses that the AI triplet could suggest, each of which represents a fruitful direction for future AI education research.


\subsection{Taking the corners of the triplet singly or together}

As an initial example, we look at one interesting and widespread set of observations/findings coming out of work on the chemistry triplet.  First, expertise in chemistry requires flexibly moving between and integrating knowledge at all three corners of the chemistry triplet:

\begin{quote}
``Trained chemists jump freely from level to level in a series of mental gymnastics.  It is eventually very hard to separate these levels.''  \cite[p. 377]{johnstone1982macro}
\end{quote}

Second, conventional practices in chemistry education would often teach concepts that spanned multiple corners of the chemistry triplet, but (a) these approaches were not often not easy for novices, as students would have to learn multiple \textit{types} of knowledge at the same time they were having to learn the new knowledge itself; and (b) these approaches did not explicitly help students recognize the different types of knowledge they were learning, or explicitly help students learn to move across knowledge types:

\begin{quote}
``[M]ost chemistry teaching is focused on the submicro–symbolic pair of the triplet and rarely helps students to build bridges to comfortably move between the three levels.''  \cite[p. 181]{talanquer2011macro}
\end{quote}




\begin{quote}
``Offering sufficient scaffolding to support students in gradually learning to operate within and across the domains in the way experts can.  So, for example, there will need to be times in teaching when the focus is on subsets of the macroscopic concepts, and how these are formally represented; and there will need to be times when the focus is on aspects of the sub-microscopic models, and the different ways these are formally represented.  There will also be times when it is vital to shift between the macroscopic and submicroscopic domains to build up the explanations of the subject....  However, ventures into the triangle should be about relating previously taught material, and should be modelled carefully by the teacher before students are asked to lead expeditions there; and such explorations should initially be undertaken with carefully structured support.''  \cite{taber2013revisiting}
\end{quote}

While undergraduate students taking AI courses generally come in with prior knowledge about programming and mathematics, it is certainly the case that teaching about any particular AI topic often introduces new knowledge and representations at all three levels of the AI triplet, as shown in Figure \ref{fig:bfs}.

Moreover, as AI education is beginning to percolate down to younger students, the need for sufficient scaffolding will only increase.  For example, a recent cognitive-interview-based study of eight middle and high school students found that students faced challenges in multiple distinct areas that correspond to facets of the AI triplet \cite{greenwald2021learning}, for instance:
\begin{itemize}
\item Mathematical:  ``Even  in  cases  where  students  demonstrated  competency 
with the necessary mathematical skills, they often struggled 
to identify connections and/or make use of those skills until 
explicitly prompted....  Once  introduced,  however, 
students generally recognized the method and were able to 
apply it to the problem.''  \cite[p. 15530]{greenwald2021learning}
\item Conceptual:  ``Students  found  difficulty  with  the  abstract 
representations characteristic in AI problems.   
Across all interviews, students needed explicit scaffolding 
in understanding how to interpret and construct a search tree.... Thus, even after 
the interviewer scaffolding enabled students to construct a 
search  tree  from  the  slider  puzzle,  students  struggled  to 
make use of the search tree representation to consider the 
depth and breadth of a problem space (both for the search 
trees they constructed and for pre-constructed exemplars). 
able to build a tree from a node to branches to new nodes, 
yet only one student was able to recognize the salience of 
tree abstractions such as branching factor and tree depth 
(albeit using colloquial language) to estimate the relative 
complexity of a problem.''  
\cite[p. 15530]{greenwald2021learning}
\item Conceptual/Computational: ``A challenge for all students interviewed, even those with 
advanced  mathematical  skills,  was  recognizing  how  a 
problem  in  the  world  could  be  made  amenable  to  the 
computational power of AI. That is to say, students needed 
support in conceiving a problem space in a way that would 
enable an AI system to solve it. Thus while some students 
in the study volunteered ways a computer program might be 
able to implement an AI solution once identified, the initial 
step  of  reconceiving  a  problem  as  an  AI  problem  was 
elusive: the broad strategies AI systems leverage to make 
predictions or to find a solution from an array of possibilities 
were unknown to students and thus unavailable resources in 
their mental models of the problem space.''  \cite[p. 15531]{greenwald2021learning}
\end{itemize}

\noindent While these interesting observations were drawn from just one small study, such detailed considerations of the \textit{types} of knowledge needed for AI teaching will be critical to innovating in AI pedagogy.  And, as suggested by the AI triplet, and by this very congruent student interview study, one potentially fruitful direction of study for AI education lies in evaluating how students are able to approach individual versus combined aspects of the AI triplet, and strategies for scaffolding for helping students build up competence along individual aspects before explicitly guiding them to think across multiple levels.

Another observation arising from the chemistry triplet is that, for experts, representations of a particular topic are often ambiguous, in that they can be interpreted in terms of multiple levels of the triplet.  This ambiguity can be challenging for students to parse: 

\begin{quote}
``e.g., at a particular moment, the teacher might be talking about molecules, but the student may be interpreting the signifiers as representing samples of substances.'' \cite{taber2013revisiting} 
\end{quote}

However, what is challenging for students is actually an important affordance for experts: 
\begin{quote}
``the affordance of this ambiguity is the potential for these symbols to allow us to shift between the macroscopic and submicroscopic levels....  An equation for a chemical reaction...can act as a bridge between the two levels by simultaneously representing both the macroscropic and submicroscopic, and aiding us in shifting between these levels in our explanations.'' \cite{taber2013revisiting}
\end{quote}

Anecdotally, while AI educational practices often present multiple and/or ambiguous representations to students, there is little focused and explicit instruction about these ambiguities and strategies students can use to learn to jump between or build bridges between levels.  For instance, again anecdotally, many standard AI exam problems (like those posted to publicly available course websites) present as either ``code'' problems or ``math'' problems or ``conceptual'' problems, without necessarily evaluating students on the ability to bridge multiple levels.  In contrast, many in-depth homework assignments or projects do require students to both conceptually describe and implement a particular AI idea.  It is worth more purposefully exploring how these kinds of bridging activities can be designed, used, and evaluated in AI education.

\subsection{The Role of Spatial Skills in AI Expertise}
\label{sec:imagination}

Taking each corner of the AI triplet independently, we can think about what kinds of cognitive competencies would feed into AI expertise, and how to support and train these competencies in our students.  That mathematical and computational proficiencies are important is no surprise, and both math and CS education are of course very mature and active fields of research.  However, it is interesting to think about what kinds of expertise are needed for the conceptual corner of the AI triplet.

In other words, when we ask students to form mental models of search trees or the surfaces of high-dimensional functions, what cognitive skills does that require?
We propose a hypothesis for AI education research that \textit{spatial skills} are a key type of proficiency that feeds into AI learning and expertise at all three corners of the AI triplet, but perhaps being especially important at the conceptual corner.

Spatial skills are cognitive skills that involve spatial perception, memory, visualization, and reasoning about spatial relationships \cite{tversky2005visuospatial}.
It is a very robust finding in the cognitive sciences that spatial skills are strongly linked to STEM learning and achievement across a variety of STEM disciplines.  For example, a very large longitudinal study found that high school spatial ability was highly predictive of occupations across STEM disciplines \cite{wai2009spatial}.  Many cognitive analyses of scientific creativity and discovery have emphasized the role of spatial visualization \cite{nersessian2008creating,miller2012insights}, and spatial skills are also linked to creativity in producing patents and publications \cite{kell2013creativity}.

It is also increasingly recognized that spatial skills are extremely important in both mathematics and CS.  For math, a recent meta-analysis found that spatial and mathematical skills are correlated in students across a wide age range, including after controlling for gender and grade level \cite{atit2021examining}, and spatial skills are prominent in accounts of expert mathematicians \cite{giaquinto_visual_2007}.  In addition, a very recent meta-analysis by Hawes and colleagues \cite{hawes2022effects} found that spatial training improves both spatial skills as well as mathematical performance in participants ranging from 3 to 20 years old.

With respect to CS learning, spatial ability is also correlated with success in early programming courses  \cite{mayer1986learning,simon2006predictors,jones2008spatial}, though of course math, logic, verbal, and other cognitive abilities also contribute \cite{wilson2001contributing}.  Spatial visualizations are also prominent in narrative accounts of expert software designers \cite{petre1999mental}.  And, recent studies have begun to show that spatial training can improve computer science learning outcomes in both high school \cite{cooper2015spatial} and college-level students \cite{bockmon2020cs1,parkinson2020effect}.

In addition, spatial skills may play a critical role in broadening participation in STEM.  A study in CS education found that spatial skills more strongly mediated the predictive relationship between soceioeconomic status and CS achievement than did computing access \cite{parker2018socioeconomic}.  And, spatial training was found to positively impact the retention of women students in an undergraduate engineering program \cite{sorby2018does}.

However, the AI triplet emphasizes that AI learning requires not just mathematical and computational proficiency, but also proficiency in thinking about conceptual knowledge at quite sophisticated levels.  We hypothesize that spatial skills are particularly important for this corner of the AI triplet, above and beyond their contributions to background abilities in computational and mathematical areas.






\section{Conclusion}


What the AI triplet adds to existing views of AI education is a way of specifying where and how different precursor competencies might feed into AI learning, and also how different aspects of an AI topic might be presented most effectively to students.  By understanding specific pathways and mechanisms of learning, we can better pinpoint where certain students may be having difficulties in learning certain types of knowledge or integrative forms of thinking, and provide additional learning supports accordingly.

The original chemistry triplet proposed by Johnstone in 1982 has elicited an enormous amount of scientific and pedagogical discourse in the field of chemistry education, including numerous refinements and re-imaginings as well as research studies about chemistry teaching and learning \cite{talanquer2011macro}.  We hope that this AI triplet does the same.

\newpage
\bibliography{references.bib}

\end{document}